\input{aipcheck}

\documentclass[
    ,final            
  ]
  {aipproc}
\usepackage{url}
\layoutstyle{8x11double}

\begin{document}

\title{The PubChemQC Project: a large chemical database from the first principle calculations}


\author{NAKATA Maho}{
  address={2F Advanced Center for Computing and Communication, RIKEN, 2-1, Hirosawa, Wako-City, Saitama, 351-0198 JAPAN}
}

\begin{abstract}
In this research, we have been constructing a large database of molecules by {\it ab initio} calculations.
Currently, we have over 1.53 million entries of 6-31G* B3LYP optimized geometries and ten excited states by 6-31+G* TDDFT calculations. To calculate molecules, we only refer the InChI (International Chemical Identifier) representation of chemical formula by the International Union of Pure and Applied Chemistry (IUPAC), thus, no reference to experimental data. These results are open to public at http://pubchemqc.riken.jp/. The molecular data have been taken from the PubChem Project (http://pubchem.ncbi.nlm.nih.gov/) which is one of the largest in the world (approximately 63 million molecules are listed) and free (public domain) database. Our final goal is, using these data, to develop a molecular search engine or molecular expert system to find molecules which have desired properties.
\end{abstract}
 
\maketitle


\section{Introduction}
One of the great achievements of quantum chemistry is that we can obtain and predict properties of molecules without experiments, even sometimes better than experiments. Nowadays, chemists cannot conduct their research without quantum chemistry program packages.

Although we have powerful a theoretical foundation \cite{Dirac}, and good implementations, still quantum chemistry only helps what the other experimental chemists do.

Our final goal is to develop a molecular search engine and expert system to find molecules which have some desired properties. Development of such system is of course difficult. Experiences and intuitions of both calculations and experiments by sophisticated chemists are indeed very important and difficult to imitate by one computer program.

Such intuitions and experiences are hard to implement, however, we believe that we can do a part of them if we restrict an imitation system to do very little further than what we can do now.
Deep learning is a very promising technique which attempt to model high-level abstractions in data, and this approach has been applied to many areas of engineering. Some very successful cases are: image \cite{GoogleCat} \cite{ImageClass} and speech recognition \cite{Speech}, and preliminary application to chemistry has also just begun \cite{Chem}.

We need a huge dataset for compounds to learn chemistry from data, however, we don't have a large database having many molecules and their properties. As far as authors' knowledge, the largest one would be NIST Chemistry WebBook \cite{NIST}. It contains over 16,000 molecules for IR spectra, 1,600 molecules for UV spectra, and other experimental data. Still, the numbers are apparently very small for deep learning; unfortunately, it costs too much for doing real experiments for many compounds.

The largest public database for molecule would be the Pubchem project \cite{pubchemproject}.
This database was assembled by the National Institutes of Health (NIH) through NIH Molecular Libraries Roadmap initiative, and provides the biological activity of small molecules. There are three projects in the Pubchem projects: Pubchem Substance, Pubchem BioAssay and Pubchem Compound, and the PubChem compound provides over 63 million of pure, standardized, and non-duplicating compounds.

Other databases are also available. The largest database from published in scientific documents would be the Chemical Abstracts Service database. It has more than 71 million organic and inorganic substances. However, it is a proprietary database, thus, hard to use for secondary use: developing a public database. The largest database using combinations would be GDB-17 \cite{GDB17}. It enumerates all possible molecules up to seventeen atoms consist of carbon, nitrogen, oxygen, sulfur, chlorine and hydrogen. Actually, there are 166 billion molecules in this database. For biochemical purpose, ChEMBL \cite{chembl} database is also popular. It was a well curated database for bio-medical and pharmacological activity. 

Our goal in this paper is developing a database for molecules from PubChem Compounds.
The reasons for employing PubChem Compounds are that the number of registered molecules is large enough (contains molecules from ChEMBL and other databases), and in the public domain. The GDB-17 enumerate molecules exhaustively, however, only small compounds are listed (typically, molecular weights are less than 300), and we do not want to restrict atoms to carbon, nitrogen, oxygen, sulfur, chlorine and hydrogen.

The calculation will complete in seventeen years, even though if we calculate ten thousand molecules per day. Our resources are limited to do calculation for thousand to ten thousand molecules per day, therefore our approach may not feasible at the moment. However, we are optimistic about it, as algorithms for quantum chemistry have been improving and computer resources increase by Moore's law.

We consider our project, which provides molecular information by quantum chemistry calculations, is beneficial as following reasons:
\begin{itemize}
\item We can provide very accurate results comparable to experiments by the {\it ab initio} calculations.
\item We can perform efficient and low-cost molecular design and search by high throughput screening.
\end{itemize}

In our PubChemQC project, the general calculation scheme for each molecule is following:
\begin{itemize}
\item Download molecular information from Pubchem compound, and we extract molecular information in International Chemical Identifier (InChI) representation \cite{inchi}. 
\item We generate initial geometry by OpenBABEL and empirical PM3 geometry optimization.
\item We further do geometry optimization by Hartree-Fock in STO-6G basis set, followed by the final geometry optimization by density functional theory (DFT) calculation employing B3LYP functional using 6-31G* basis set.
\item Using the final molecular geometry, we calculate ten lowest excited states by time depended density functional theory (TDDFT) calculation using 6-31+G* basis set.
\item Upload the results with input files to http://pubchemqc.riken.jp/.
\end{itemize}


\section{How to proceed with the calculation}
In the following subsections, we show the outline how PubChemQC project has been doing.

\subsection{Information stored in PubChem Compound}
We acquired information of the molecule from PubChem project\cite{pubchemsdf}.
There were 63,159,671 molecules in 2015/2/4 \cite{molcount} and this number has been updated daily.
The molecules at PubChem Project are provided in SDF (structure-data file). There are approximately 3,000 SDF files and each file contain almost 25,000 molecules. All the molecules are almost unique and some molecules are obsoleted and removed. A SDF of PubChem Compound contains
\begin{itemize}
\item Three dimensional structures without hydrogen via PubChem 3D project\cite{PubChem3D}.
\item IUPAC names. 
\item InChI and SMILES representations of molecule
\item Molecular weights,
\end{itemize}
etc. We actually use the InChI representation of molecules and the molecular weights. Other information are not referenced.

\subsection{InChI representation of molecule}
The IUPAC International Chemical Identifier is a textual identifier for chemical substances, designed to provide a standard and human-readable way to encode molecular information and to facilitate the search for such information in databases and on the web (from Wikipedia).

Ethanol is represented as follows:
\begin{verbatim}
InChI=1S/C2H6O/c1-2-3/h3H,2H2,1H3
\end{verbatim}
L-ascorbic is represented as follows:
\begin{verbatim}
InChI=1S/C6H8O6/c7-1-2(8)5-3(9)4(10)6(11)12-5/h2,5,7-8,10-11H,1H2/t2-,5+/m0/s1
\end{verbatim}
.

\subsection{Details of quantum chemical calculation}
First, we made a file containing compound identification number (CID) with InChI representation and molecular weight in each line of PubChem Compound. This file was approximately 7.8G bytes and contained 51,520,346 lines.

Then, we sorted by molecular weight in ascending order so that earlier lines contain lighter molecules; our calculation started from hydrogen and will end in very heavy molecule. In Figure 1, we show how molecular weight distributes in PubChem Compound. Interestingly, the sorted CID does not show as log scale. It shows that the number of molecules known to us scale as linearly whereas we can just enumerate astronomical number of molecules by combinatrics. Roughly saying, we can calculate lighter molecules much faster and much easier than heavier molecules. For heavier molecules, we should consider many conformers, better basis sets and relativistic effects, however, these points are out of scope at this moment.

\begin{figure}
\includegraphics[height=.2\textheight]{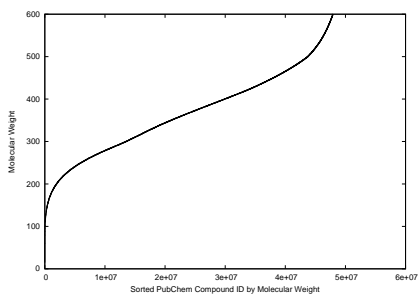}
\caption{Molecular weight distribution of PubChem Compounds.}
\end{figure}

There are some compounds not suitable for calculation: a mixture like ionic salts ($\rm Ni^{2+} SO_4^{2-}$). These systems are omitted. Unfortunately, molecules contains $\eta^5$ bonding (e.g. Ferrocene) are also not calculated, due to problem in InChI representation. Moreover, we only calculate molecules containing H, He, Li, Be, B, C, N, O, F, Ne, Na, Mg, Al, Si, P, S, Cl, Ar, K, Ca, Sc, Ti, V, Cr, Mn, Fe, Co, Ni, Cu and Zn. This is due to limitation of 6-31G* basis set. 

The initial geometry for a molecule was made by OpenBABEL with ``--gen3d -addH'' options from InChI representation.
Then, the first geometry optimization calculation is performed by PM3 method, resultant geometry was further optimized
by Hartree-Fock method using STO-6G basis set. Finally, we re-optimized by density functional theory of B3LYP functional
using 6-31G* basis set. The final optimization process has actually three steps; the first step
was done by FireFly \cite{FireFly}, then the second step was done by GAMESS \cite{GAMESS}, the third step
was again done by GAMESS, since calculation time by FireFly was substantially faster, and slightly less accurate.
And why we do optimization by GAMESS twice is to make sure that the final optimization was really final; only one geometry search should be done.

Subsequent excited state calculation was done by TDDFT with 6-31+G* basis set employing same geometry as the final optimization of previous calculation.

Final results and input files are uploaded at http://pubchemqc.riken.jp/.

All calculations are done on RICC supercomputer (Intel Xeon 5570 (2.93GHz) x 2, 1024 nodes) at RIKEN,
Quest supercomputer (Intel Core2 CPU L7400 @ 1.50GHz, 700 nodes) at RIKEN, and Oakleaf-FX supercomputer (Fujitsu PRIMEHPC FX10, SPARC64 IX fx@1.848GHz) at the University of Tokyo.
With above computational resources, it is possible to calculate thousand to ten thousand molecules per day.

\subsection{Issues and discussion}
There are some issues on calculations.

The first one is issue on theoretical parameters: which basis function, calculation method should be used. Of course,
better basis is better (e.g., cc-pVQZ) and better method is better (e.g., CCSD(T)). However, if we employ such choices, calculation would take extremely long time. We believe our choice, 6-31G* basis set and DFT with B3LYP, are fairly good compromise
between quality of results and resources we have. Usually, properties obtained by B3LYP functional are comparable to MP2 calculations. In some cases, we may need to employ some more expensive methods. We just left such molecules as calculation failed.

The second one is that the resultant molecular structures are not necessarily global minimum energy ones, as initial structure issued by OpenBABEL from InChI, each optimization process may not appropriate ones. Moreover, there should be conformers and secondary structures. If we want to make sure that obtained geometries are really global minimum of the molecule, we need to perform global geometry search like GRRM method \cite{GRRM}, which will take extremely very long time.
Nevertheless, we expect current resultant structures are reasonable, not so funny.

The third one is limitation of InChI representation of molecule. A clear example for InChI representation
does not work is Ferrocene. We may need to extend InChI representation. Unfortunately, any representation
has its limitation. Most general one may be store in three dimensional coordinates of nucleus.
Of course, we need to compromise between usefulness and having many corner cases.

The fourth one is which molecules to be calculated. The GDB-17 has astronomical number of molecules \cite{GDB-17}, even,
these molecules consist of at most seventeen of atoms. Which molecules considered to be important is a difficult problem.
One possible answer might be ``All important molecules are listed in PubChem or in CAS database, but there should be more''.

\section{Summary and outlook}
We described research outline of the PubChemQC project, we have been constructing
a database listed in PubChem Compound by NIH by {\it ab initio} calculations. We only refer chemical formula in InChI of
PubChem compound, therefore, no reference to experimental data. Currently, optimized geometries, ground state energies and excited state energies of 15.3 million molecules are calculated by GAMESS, FireFly and OpenBABEL, and uploaded to http://pubchem.riken.jp/. We employed calculation method as density functional theory with B3LYP functional and 6-31G* and 6-31+G* basis set.

Since the results are presented as input and output files, the reproducible results and reuse Is easy.
Within the current computational resources, it is possible to calculate thousand to ten thousand molecules per day.

As the future plan, we will construct web site, for compound search, and calculation of other properties like
vibrating structure, NMR chemical shift, and structure optimization in the excited state and with solvent effects.

\begin{theacknowledgments}
We are very grateful for Motoyoshi KUROKAWA and Ryutaro HIMENO, and Masashi Horikoshi for kind advices, maintenance of computers, and providing resources. The calculations were performed by using the RIKEN Integrated Cluster of Clusters (RICC) facility and the research is partially supported by the Initiative on Promotion of Supercomputing for Young or Women Researchers, Supercomputing Division, Information Technology Center, The University of Tokyo.
\end{theacknowledgments}

\bibliographystyle{aipproc}

\end{document}